\documentstyle[aps,prb,twocolumn]{revtex}
\begin{document}
\title{\hspace{1cm} Physics Letters A 225 (1997) 167-169\\ \vspace{1cm}
Dipole-dipole interaction of Josephson diamagnetic moments}
\author{Sergei A. Sergeenkov}
\address{Bogoliubov Laboratory of Theoretical Physics,
Joint Institute for Nuclear Research,\\ 141980 Dubna, Moscow region, Russia}
\address{
\centering{
\medskip\em
\begin{minipage}{14cm}
{}~~~
The role of dipole-dipole interaction between Josephson diamagnetic moments
is considered within a model system of two clusters (each cluster contains
three weakly connected superconducting grains). The sign of the resulting
critical current is shown to depend on the orientation between clusters,
allowing for both $0$ and $\pi$ type junctions behavior. The possibility of
the experimental verification of the model predictions is discussed.
{}~\\
\medskip
{}~
\end{minipage}
}}
\maketitle
\narrowtext

 A weak-link structure (both intrinsic and extrinsic) of high-$T_c$
 superconductors (HTS) is known to play a rather crucial role in
 understanding many unusual and anomalous physical phenomena in these
 materials (see, e.g., [1-7]). In particular, the "fishtail" anomaly of
magnetization in oxygen-deficient crystals is argued [4-7] to originate
 from intrinsic (atomic scale) weak links, while the spontaneous orbital
 magnetic moments induced by grain boundary weak links with the so-called
 "$\pi $ junctions" (see, e.g., [8]) is believed to be responsible for
 the "paramagnetic Meissner effect" [9,10] in granular superconductors.
Furthermore, to probe into the symmetry of the pairing mechanism in HTS,
the Josephson interference experiments based on the assumption that $\pi$
junctions are created between $s$-wave and $d$-wave superconductors [11]
or between different grains of the $d$-wave superconductor itself [12]
have been conducted (see, e.g., [13] for discussion and further references
therein).

 In the present paper, we would like to draw a particular attention to the
 importance of the dipole-dipole interaction between the above-mentioned
 orbital magnetic moments. Due to the vector character of this interaction,
 the sign of the resulting Josephson critical current will depend on the
 orientation between these diamagnetic moments, giving rise to either $0$ or
 $\pi$ type junction behavior. It is shown that for large enough grains and/or
 small enough distance between clusters, the dipole energy between clusters
 may overcome the direct Josephson coupling between grains within a single
 cluster, allowing for manifestation of long-range correlation effects in
 granular superconductors.

  Let us consider a system of two identical clusters of superconducting
  grains. Each cluster contains three weakly connected grains (which is a
 minimal number needed to create a current loop and the corresponding
 non-zero diamagnetic moment $\vec \mu $, see below). Between adjacent grains
 in each cluster, there is a Josephson-like coupling with the energy
 $J_{ij}=J_{ji}$. The dipole-dipole interaction between these two clusters
 can be presented in the form
\begin{equation}
{\cal H}_{dip}=\sum_{\alpha \beta }^{}V_{\alpha \beta }(\vec R)\mu ^{\alpha }
\mu ^{\beta },
\end{equation}
where
\begin{equation}
V_{\alpha \beta }(\vec R)=\frac{\mu _0}{R^3}\left( \delta _{\alpha \beta}-
\frac{R_{\alpha }R_{\beta }}{R^2}\right )
\end{equation}
and
\begin{equation}
 \mu ^{\alpha }=\frac{2e}{\hbar }\sum_{i=1}^{3}J_{i,i+1}\sigma _{i,i+1}^{
 \alpha }\sin \phi _{i,i+1}.
 \end{equation}
Here, $R$ is the distance between clusters, $\vec \sigma _{ij}=\vec r_i\times
 \vec r_j$ is the (oriented) projected area for each cluster, $\phi _{ij}=
\phi _i-\phi _j$ is the phase difference between adjacent grains, and
$\mu _0=4\pi \times 10^{-7}H/m$. Hereafter, $\{ \alpha ,\beta \}=x,y,z$; and
$\{ i,j\}=1,2,3.$

In view of the obvious constraint, $\phi _{12}+\phi _{23}+\phi _{31}=0$,
our consideration can be substantially simplified by introducing a
"collective variable" $\phi $, namely
\begin{equation}
\phi _{12}=-\phi _{31}\equiv \phi ,\qquad \phi _{23}=0
\end{equation}
As a result, the dipole-dipole interaction energy takes on a simple form
\begin{equation}
 {\cal H}_{dip} =D\sin ^2{\phi },
\end{equation}
where
\begin{equation}
D=\left( \frac{2e}{\hbar}\right )^2
\sum_{\alpha \beta }^{}V_{\alpha \beta }(\vec R)\left( J\sigma \right )^
{\alpha }\left( J\sigma \right )^{\beta },
\end{equation}
with
\begin{equation}
\left( J\sigma \right )^{\alpha }\equiv J_{12}\sigma _{12}^{\alpha }+
J_{13}\sigma _{13}^{\alpha }.
\end{equation}
To estimate the significance of the above-considered dipole-dipole energy,
we have to compare it with the Josephson coupling energy between grains
within a single cluster. The latter contribution for three adjacent grains
(forming a cluster and allowing for a non-zero current loop) gives
\begin{equation}
{\cal H}_J=-\sum_{i=1}^{3}J_{i,i+1}\cos \phi _{i,i+1},
\end{equation}
or equivalently, in terms of the "collective variables" (see Eq.(4))
\begin{equation}
{\cal H}_J=-J_{23}-(J_{12}+J_{13})\cos \phi .
\end{equation}

Thus, the resulting Josephson current in our system of the two coupled
clusters, defined as $I(\phi )=(2e/\hbar )(\partial {\cal H}_{tot}/\partial
\phi )$ with the total energy ${\cal H}_{tot}=2{\cal H}_J+{\cal H}_{dip}$
(notice that there is no direct Josephson interaction between clusters,
they are coupled only via the dipole-dipole interaction) reads
\begin{equation}
I(\phi )=2I_J^c\sin \phi +I_D^c\sin 2\phi ,
\end{equation}
where
\begin{equation}
I_J^c=\frac{2e}{\hbar }(J_{12}+J_{13}),
\end{equation}
and
\begin{equation}
I_D^c=\frac{2e}{\hbar }D.
\end{equation}
It is interesting to mention that a similar to Eq.(10) form of the
"nonsinusoidal" current-phase relationship has been recently discussed by
Yip [13] who investigated the Josephson coupling involving unconventional
superconductors beyond the tunnel-junction limit.

Let us find out now when dipole-dipole interaction between two clusters
may become comparable with (or even exceed) the direct Josephson coupling
between grains within the same cluster. In view of Eq.(10), this will
happen when $D$ becomes equal to (or larger than) $2(J_{12}+J_{13})\approx
4J$ which in turn is possible either for small enough distance between
clusters $R$ or for large enough grain size $r_g\approx \sqrt{\sigma /\pi }$.
Taking $J/k_B\approx 90K$ for the maximum (zero-temperature) Josephson energy
in $YBCO$ materials, and assuming (roughly) $R\approx r_g$, we get
$r_g\approx 10\mu m$ for the minimal grain size needed to observe the effects
due to the dipole-dipole interaction between diamagnetic moments in granular
HTS.

There is however another possibility to observe these effects which does not
require the above-mentioned restrictions (small $R$ and/or large $r_g$).
Indeed, in view of Eq.(11), when grains $1$ and $2$ form a $0$ junction (with
$J_{12}=J$) and at the same time grains $1$ and $3$ produce a $\pi$ junction
(with $J_{13}=-J$) within the same cluster, the direct Josephson contributions
cancel each other so that $I_J^c\equiv 0$, and the resulting critical current
is completely defined by its dipole part only, which in this case reads
$$
I_D^c=\left( \frac{2e}{\hbar}\right )^3J^2
\sum_{\alpha \beta }^{}V_{\alpha \beta }(\vec R)\Delta \sigma ^{\alpha }
\Delta \sigma ^{\beta },
$$
with $\Delta \sigma ^{\alpha }\equiv \sigma _{12}^{\alpha }-\sigma _{13}^
{\alpha }$.
Moreover, due to the orientational nature of the dipole-dipole interaction,
the induced critical current $I_D^c$ may exhibit properties of either
$0$ or $\pi$ junctions, depending on the sign of the interaction potential
$V_{\alpha \beta }(\vec R)$. In particular, as is seen from Eq.(2), the
non-diagonal part of $V_{\alpha \beta }(\vec R)$ is responsible for creation
of $\pi$ type junctions with $I_D^c<0$. It would be very interesting to
try to observe the predicted behavior experimentally, using perhaps some
artificially prepared systems of superconducting grains.

In summary, a system of two clusters (of weakly connected superconducting
grains) coupled via dipole-dipole interaction between their diamagnetic
moments was considered. For large enough grains (or small enough distance
between clusters), the dipole energy between clusters was found to compare
with the direct Josephson coupling between grains within a single cluster.
The sign of the critical current, related to the dipole energy, was shown
to depend on the mutual orientation of the clusters, varying from $0$ to
$\pi$ type junction behavior.

\end{document}